\documentclass[aps,twocolumn,prx,preprintnumbers,showpacs,amsmath,amssymb,superscriptaddress,pra]{revtex4-1}
\usepackage{epsfig}

\newcommand{\ii}{\mathrm{i}}

\begin{document}
\title[]{$GW$100: a plane wave perspective for small molecules}

\author{Emanuele Maggio}
\affiliation{University of Vienna, Faculty of Physics and Center for
Computational Materials Science, Sensengasse 8/12, A-1090 Vienna, Austria}

\author{Peitao Liu}
\affiliation{University of Vienna, Faculty of Physics and Center for
Computational Materials Science, Sensengasse 8/12, A-1090 Vienna, Austria}
\affiliation{Institute of Metal Research, Chinese Academy of Sciences, Shenyang 110016, China}

\author{Michiel J. {van Setten}}
\affiliation{Nanoscopic Physics, Institute of Condensed Matter and Nanosciences, Universit\'{e} Catholique de Louvain, 1348 Louvain-la-Neuve, Belgium}

\author{Georg~Kresse}
\email{georg.kresse@univie.ac.at}
\affiliation{University of Vienna, Faculty of Physics and Center for
Computational Materials Science, Sensengasse 8/12, A-1090 Vienna, Austria}

\date{\today }

\begin{abstract}

In a recent work, van Setten and coworkers have presented a carefully converged
$G_0W_0$ study of 100 closed shell molecules [J. Chem. Theory Comput. \textbf{11}, 5665 (2015)]. For two different
codes they found excellent agreement to within few 10 meV
if identical Gaussian basis sets were used.
We inspect the same set of molecules using the projector augmented wave method
and the Vienna ab initio simulation package (VASP).
For the ionization potential, the basis set extrapolated plane wave results agree very well with the
Gaussian basis sets, often reaching better than 50 meV agreement.
In order to achieve this agreement, we correct for finite basis set
errors as well as errors introduced by periodically repeated images.
For electron affinities below the vacuum level differences between
 Gaussian basis sets and VASP are slightly larger. We attribute this
to larger basis set extrapolation errors for the Gaussian basis sets.
For quasi particle (QP) resonances above the vacuum level, differences between VASP
and Gaussian basis sets are, however, found to be substantial.
This is tentatively explained by  insufficient basis set convergence of the Gaussian type orbital calculations
as exemplified for selected test cases.

\end{abstract}

\maketitle

\section{Introduction}

The $GW$ approximation suggested by Lars Hedin \cite{Hedin_GW_1965} has a long history in solid state physics.
First practical applications were already published in the 1980s
by Hanke and coworkers soon followed by
the often quoted study of Hybertsen and Louie \cite{Hanke1979,strinati1980,strinati1982,Hybertsen_GW_1986}.
For solids, it is generally found that even the simplest approximation $G_0W_0$ yields
reasonably accurate quasiparticle (QP) energies and band gaps in good agreement with experiment \cite{Schilfgaarde_QPGW_2006,Shishkin2007,Fuchs2007}.
The results often improve if the Green's function is iterated
to self-consistency, either updating the QP-energies only
or even the one-electron orbitals \cite{Faleev_QPGW_2004,Schilfgaarde_QPGW_2006,Shishkin2007,Caruso2012a,Caruso2013a,Kaplan2015,Kaplan2016,Lischner2014a,Koval2014}. Applications
of the $GW$ approximation to molecules, however, have been comparatively rare, since
codes based on local orbitals, which are by construction particularly
well suited to treat molecules, did not incorporate the $GW$ approximation
until recently. This has changed,
with many local basis set codes, such as FHI-aims,  MOLGW, Turbomole, and CP2K now supporting  $GW$
calculations \cite{Ren2012,Blase_GW_molecules_2011,Bruneval2012,vansetten13,Wilhelm2016,Foerster2011,Ke2011}.
Also, efficient plane wave codes using a Sternheimer approach, such as ABINIT and West \cite{abinitsternh,Govoni_West_2015}, 
are becoming available.
As for solids, carefully converged QP calculations are, however, still comparatively scarce \cite{Klimes_predictiveGW_2014}.

To fill this gap, Bruneval recently performed systematic studies for
about 30 molecules \cite{Bruneval2013}.
van Setten and coworkers went one step further and evaluated basis set extrapolated
$GW$ QP energies for 100 closed shell molecules using several codes \cite{Setten_GW100_2015}.
They found that the $GW$ QP energies of  the highest
occupied orbital (HOMO) and lowest unoccupied orbital (LUMO) of
two local basis set codes, FHI-aims and Turbomole, virtually agree, {\em if identical  basis sets are used}.
In many respects this is not astonishing, since two codes ought
to yield the same results, if the computational parameters are identical. The two codes are, however,
technically quite different. For instance, they introduce auxiliary basis sets
to avoid storing the two-electron four orbital integrals. Furthermore, FHI-aims 
uses a numerical representation of the Gaussians and calculates
the self-energy along the imaginary axis (Wick rotation) requiring an
analytic continuation to the real axis. All these factors can introduce small uncertainties.
Clearly, the study impressively demonstrates that all these intricacies
are well under control, and technically well converged results can be obtained using both
codes.

The paper by van Setten {\em et al.}\cite{Setten_GW100_2015} also reports results using the $GW$ Berkeley plane wave
code \cite{Deslippe2012}. Although agreement of that code with experiment is very good if
the plasmon-pole model is used, comparison of the fully frequency dependent
$G_0W_0$ HOMO and LUMO with Gaussian basis set results is
less satisfactory. For the considered molecules, the mean absolute difference between
Gaussian type orbitals (GTO) and plane waves (PWs) is about 200 meV for the HOMO.
We note on passing that the agreement between GTO and other plane wave studies is
seemingly superior \cite{Govoni_West_2015,abinitsternh}, although, this could be related to the fact that
these studies only considered a subset of the $GW$100 set.
The disagreement between the Berkeley $GW$ PW code and GTO codes is certainly slightly disconcerting,
since it puts decades of studies using PW based
$GW$ calculations into question. Remarkably, on the level of DFT, the reported one-electron energies
of  the HOMO agree to within few 10 meV. So how can one understand
the much larger discrepancies  for $GW$ QP  energies?

A partial answer is given by the observation that QP energies converge very slowly with respect
to the basis set size, as well established for Gaussian type orbitals \cite{Bruneval2013,Setten_GW100_2015,Wilhelm2016}.
van Setten {\em et al.} obtained basis set converged  QP energies  by extrapolating
against the basis set size or against $1/C_n^3$, where $C_n$ is the basis set
cardinal number \cite{Setten_GW100_2015}. Extrapolation  was based on def2-SVP, def2-TZVP and def2-QZVP, but even though
def2-QZVP constitutes a fairly complete set, the extrapolated values can differ
by more than 300 meV from the values at the largest considered basis set.
Astonishingly, the reported PW results were not extrapolated to the basis set
limit, although a recent work of Klimes {\em et al.}  shows that the
$GW$ QP energies converge like one over the number of plane waves \cite{Klimes_predictiveGW_2014}
and this behaviour is also confirmed by purely analytical arguments \cite{Schindlmayr2013}.
Early evidence of this slow convergence using PWs exist aplenty \cite{shih2010,friedrich2011,Bruneval2008}.
In view of this slow convergence,
a brute force approach to predict QP energies seems elusive considering that
most codes scale cubically with respect to the number of  basis functions.
The present work tries to rectify this issue by reporting QP energies
using the plane wave code VASP, carefully correcting for basis set incompleteness errors, as detailed in section \ref{sec:theory}.

Another point that we briefly mention in section \ref{sec:results} is
 that the calculation of the poles of  the $G_0W_0$ Green's function can
 be unphysical, if the initial Green's function yields too small excitation energies.
 In this case, first linearizing the $G_0W_0$  self-energy and then determining the poles of the Green's function
 yields more robust QP energies. We, finally, finish with discussions
and our conclusions.

\section{Theory and Computational method}
\label{sec:theory}
\subsection{Theory}

$GW$ is a well established perturbative approach to calculate QP energies \cite{Hedin_GW_1965}.
In the $GW$ approximation, one initiates the calculations using a groundstate
DFT calculation to obtain the DFT one-electron orbitals $\phi_n$ and the corresponding one-electron energies $\epsilon_n$.

The first step in a $GW$ calculation is to determine the DFT Green's functions,
\begin{equation}
 G_0({\bf r}', {\bf r}, \omega) =
 \sum_n   \frac{\phi_n({\bf r}') \phi^*_n({\bf r})}{ \omega - \epsilon_n - i \eta \,{\rm sign}( \mu-\epsilon_n)},
\end{equation}
where $\mu$ is the chemical potential of the electrons, and $\eta$ a positive
infinitesimal.
From the Green's function the independent particle polarizability
\begin{equation}
 \label{equ:chi}
 \chi({\bf r}, {\bf r}', t ) = -\ii G_0({\bf r}, {\bf r}', t)G_0({\bf r}', {\bf r}, -t),
\end{equation}
and the corresponding screened interaction
\begin{equation}
 W({\bf r}, {\bf r}', \omega)= v({\bf r}, {\bf r}') +v({\bf r}, {\bf s}) \chi({\bf s}, {\bf s}', \omega) W({\bf s}', {\bf r}', \omega)
\end{equation}
can be determined. Here $v$ is the Coulomb kernel, and integration over repeated spatial coordinates ($\bf s$ and ${\bf s}'$) is assumed.
Furthermore, the  Green's functions and polarizabilities in frequency
and time domain are related by a Fourier transformation. The final step is to calculate the interacting Green's function
\begin{equation}
 \label{equ:Gsigma}
 G({\bf r}, {\bf r}', \omega)= \frac{1}{\omega - T - V^{\rm H}({\bf r}) \delta({\bf r}-{\bf r}')- \Sigma({\bf r}, {\bf r}', \omega)},
 \end{equation}
where  $T$ is the kinetic energy operator, $V^{\rm H}$ is the Hartree-potential, and  $\Sigma({\bf r}, {\bf r}',t)$ is the self-energy in the $GW$ approximation:
\begin{equation}
 \label{equ:sigmaGW}
 \Sigma({\bf r}, {\bf r}',t) = \ii G_0({\bf r}, {\bf r}',t) W({\bf r}, {\bf r}',t) .
\end{equation}
The poles of the Green's function then determine the QP energies.
In principle, this cycle can be continued by evaluating $\chi$ in step (\ref{equ:chi}) using the updated Green's function
and iterated to self-consistency. It is also
possible to obtain partial self-consistency, for instance, by calculating $W$
once and forever using the DFT orbitals and one electron
energies and iterating only the Green's function until it is self-consistent
[{\em i.e.} iterating only Eqs. (\ref{equ:Gsigma}) and  (\ref{equ:sigmaGW})].

The most common approximation is, however, the $G_0W_0$ approximation
e.g. used by Hybertsen and Louie \cite{Hybertsen_GW_1986}.
Instead of the poles of the Green's function, this approximation calculates
the nodes of the denominator in Eq. (\ref{equ:Gsigma})
 \begin{equation}
 \label{equ:QPfull}
  E^{\rm QP}_n=\text{Re}\left[\langle \phi_n |T  + V^H({\bf r})+ \Sigma( E^{\rm QP}_n) | \phi_n \rangle\right]
\end{equation}
in the basis of the DFT orbitals.
Since this involves only the diagonal elements of the self-energy, solutions of this
equation are
cheaper to determine than poles of the fully interacting Green's function.
Obviously this
is a good approximation, if the self-energy is diagonally dominant in the
basis of the DFT orbitals. As already pointed out by Hybertsen and Louie
this is generally the case, although there is some evidence
that iterating the DFT orbitals is important \cite{Faleev_QPGW_2004,Schilfgaarde_QPGW_2006,Bruneval_CHOSEX_2006}.
This is particularly so for atoms or molecules, since the KS potential
and, as a result, the KS orbitals do not decay properly at large distances from the molecule.

The solutions obtained by solving Eq. (\ref{equ:QPfull}) are labeled as  $G_0W_0$ in the present work.
Furthermore, a commonly used approximation is to linearize the the energy dependence in the self-energy in Eq. (\ref{equ:QPfull}) 
at the DFT one-electron energy and determine the nodes of the linearized equation.
This yields the following approximate position for the nodes \cite{Hybertsen_GW_1986,KresseGWa}:
\begin{equation}
 \label{equ:QPlinear}
 E^{\rm QP}_n -\epsilon^{\rm DFT}_n =
 Z_n \text{Re}\left[\langle \phi_n |T  + V^{\rm H}({\bf r})+ \Sigma( \epsilon^{\rm DFT}_n) -\epsilon^{\rm DFT}_n| \phi_n \rangle\right]
\end{equation}
were $Z_n$ is related to the derivative of the self-energy at $\epsilon^{\rm DFT}_n$
\[
 \label{equ:Z}
 Z_n = \Big(1-\left. \frac{\partial\text{Re}[\langle \phi_n|\Sigma(  \omega)|\phi_n \rangle]}{\partial\omega} \right|_ {\epsilon^{\rm DFT}_n}\Big)^{-1}.
\]
The correlation factor $Z_n$ can be also related to the amplitude of the corresponding QP peak and is a measure
of the degree of correlation. For the HOMO and LUMO of molecules, $Z$ is commonly between 0.7-0.9,
corresponding to a low to very low degree of correlation. Solutions
of the linearized equations will be labeled as  lin-$G_0W_0$ in the present work, and 
the first derivative is evaluated using central difference with the step size of $\Delta =\pm 0.1$~eV.

\subsection{Technical details}

As in the $GW$100 paper of van Setten {\em et al.} \cite{Setten_GW100_2015}, we use the PBE functional for the DFT starting point.
However, all calculations  include scalar relativistic effects, in contrast to
the calculations of van Setten {\em et al.} that are based on
non-relativistic potentials.
The potentials used in the present work are the $GW$ potentials distributed with the latest
release of VASP (vasp.5.4), and we followed the recommendations in the VASP manual
on which version to use. Generally this means
that lower lying semi-core states were not correlated in the calculations,
except for the alkali and alkali-earth metals, as well as Ti and Ga.
For He, we found issues with the originally distributed potential. The
{\tt He\_GW} potential failed to converge in DFT calculations when
the plane wave cutoff was increased, because a ghost state was introduced as
the basis set size increased. The potential was slightly modified to remove this problem
and will be distributed with the next release.
Furthermore, for boron to fluorine the potentials {\tt B$\_$GW$\_$new}, ...,
{\tt F$\_$GW$\_$new} were used (also already distributed with vasp.5.4). 
These potentials include $d$ partial
waves, whereas the standard $GW$ potentials choose the $d$ potential as the local potential.

The potentials used in this work are not the most accurate $GW$ potentials
yet available for VASP. Specifically, we have recently shown that norm-conserving (NC)
 $GW$ potentials are necessary to predict very accurate QP energies for $3d$, $4d$ and $5d$ elements~\cite{Klimes_predictiveGW_2014}
 with the NC potentials generally increasing the QP binding energies. In our experience, such highly accurate
potentials are, however, not required in the present case for the following reasons.
For $s$ and $p$ elements the standard potentials conserve the norm very well
to within about 70~\%, often even 90~\%.  Furthermore, errors introduced by violating the norm-conservation
can only occur at very high scattering energies, since the standard $GW$-PAW potentials
predict the scattering properties correctly up to about 400 eV. Beyond that energy, the PAW projectors become
incomplete. For the elements considered here, we expect that the combination of
these two effects means that the results for the HOMO and LUMO will be accurate
even though we do not use NC potentials. The only exceptions are copper,
neon, fluorine, oxygen and possibly nitrogen. These elements possess strongly localized
$3d$ and $2p$ orbitals.  We will return to this point later.

In the calculations presented here we calculate the Green's function, the
screened interaction $W$ as well as the self-energy in imaginary
time and frequency. This has several advantages compared to the full real
frequency implementation also available in VASP.
The fully frequency dependent version along the real axis requires at least
100, but for molecules with their sharp resonances often even several hundred frequency points
to converge.
Since the boxes considered in this work are quite large, we also need several  thousands of plane waves
to describe  the frequency dependent screened
interaction and Green's function accurately. This  becomes very quickly prohibitive. In the
imaginary frequency, on the other hand, only relatively few frequency points are required.
In the calculations presented here, 16 frequency points and the time and frequency grids
discussed by Kaltak {\em  et al.} are used \cite{Kaltak2014,Kaltak_SiLowScaling_2014}.  
These 16 points were found to be sufficient to converge the
QP energies of the HOMO and LUMO to about 10 meV \cite{Liu2016}.
The downside of working in the imaginary frequency domain is that the
results along the imaginary frequency axis need to be continued to
the real axis. This was done using a (16 point) Pad\'e fit following
Thiele's  reciprocal difference method~ based on continued fractions \cite{pade1975}. We note that the reported
FHI-aims results in Ref. \onlinecite{Setten_GW100_2015} were--- with few problematic exceptions ---also obtained using
16 parameter Pad\'e fits. These exceptions are BN, O$_3$, BeO, MgO and CuCN where
many more points were required. For the other molecules, the 16 parameter Pad\'e fits
yielded excellent agreement with Turbomole, which  calculates the exact $GW$ self-energy along the real axis.
Details of our implementation are reported elsewhere \cite{Liu2016}.

The other crucial issues are basis set extrapolation and convergence with
respect to the box size. To obtain basis set converged results,
we used a relatively small box, but one that still faithfully reproduces the character
of the HOMO and LUMO.
For this box, we performed calculations for
the default cutoff as specified by the VASP potentials, and calculations for three
additional plane wave cutoffs, with the largest calculation corresponding
to twice the number of plane waves used in the default setup.
These four data points are fitted assuming that the QP energies
as a function of the number of plane waves $N_{\rm pw}$ converge like
\begin{equation}
\label{equ:basisextra}
 E^{\rm QP}(N_{\rm pw})= E^{\rm QP}(\infty) + \frac{C}{  N_{\rm pw}},
\end{equation}
where $N_{\rm pw}$ is the number of plane waves in the basis set \cite{Harl_nobelgas_2008,Shepherd_convergence_2012,Gulans_RPA_layered_2012,Klimes_predictiveGW_2014}.
A four point fit and a two point fit with the largest and smallest PW basis set
yielded a maximum difference of 10 meV in the QP energies. To illustrate that the basis set
dependence is indeed following a $1/{  N_{\rm pw}}$
behavior to great accuracy, we will show data for selected molecules in Sec. \ref{sec:convergencePW}.
The only subtlety impeding an accurate and automatic extrapolation is the use of the Pad\'e fit.
The slope of the self-energy can vary somewhat between different calculations
causing some variations in the predicted QP energies. Extrapolation from these
``noisy'' data is difficult and error prone. To circumvent this issue, we perform
the extrapolation for the self-energy evaluated at the DFT one-electron
energies, specifically  on
$\Delta E = \text{Re}\left[\langle \phi |T+V^{\rm H}+\Sigma(\epsilon^{\rm DFT})| \phi \rangle\right]- \epsilon^{\rm DFT}$
instead of $\Delta E = E^{\rm QP}- \epsilon^{\rm DFT}$,
and scale the correction by the  $Z$-factor at the smallest, {\em i.e.} default, PW cutoff.
In this way, we neglect variations of the $Z$-factor  between different
basis sets, but these variations are small and dominated by noise.

A few final comments are in place here. In the calculations presented
herein, we calculate {\em all} orbitals spanned by the PW basis set.
This implies that the number of orbitals also increases as the number of
plane waves increases. Second, the kinetic energy  cutoff for
the response function ({\tt ENCUTGW} in VASP) is set to 2/3 of the
cutoff used for the plane wave basis of the orbitals ({\tt ENCUT} in VASP).
Whenever the PW cutoff for the orbitals is increased, the
PW cutoff for the basis set of the response function is increased
accordingly. This means that a single parameter, the PW cutoff for
the orbitals ({\tt ENCUT}), entirely controls the accuracy of the calculations (at least
with respect to the basis sets). Since all the intermediate
control parameters are set automatically by VASP, and
since the QP energy corrections  converge like one over  the number of plane waves and orbitals \cite{Klimes_predictiveGW_2014},
extrapolation to the infinite basis set limit is straightforward and robust.

Let us now comment on the second point, convergence with respect to the cell size.
In plane wave codes,
it is common practice to truncate the Coulomb kernel at a certain distance $r_c$,
say half the box size, so that the periodically repeated orbitals
can not screen the central atom. The downside of this
approach is that it modifies the Coulomb kernel to become\cite{Rozzi_Coulombcutoff_2006}
\[
 \frac{4 \pi e^2}{|{\bf g}|^2} (1-\cos ( |{\bf g}| r_c)),
\]
where $\bf g$ is a plane wave vector.
Obviously, this modifies the Coulomb kernel  at large reciprocal lattice vectors $\bf g$.
In test calculations we found that this spoils the previously mentioned
basis set extrapolation (\ref{equ:basisextra}): as one increases the plane
wave cutoff, one moves through maxima and minima of the truncated Coulomb kernel,
causing superimposed oscillations in the QP energies. Basis set
extrapolation becomes then uncontrolled.
To deal with the repeated images, we instead resort to the standard
trick used in periodic codes: $k \cdot p$ perturbation theory \cite{Baldereschi_dielectric_properties_78}.
We calculate the first order change of the orbitals with respect to
$k$ \cite{Gajdos_optical_properties_2006}, and accordingly the head and wings of the polarizability and a correction
to the ${\bf g} \to 0$ component of the self-energy. This term
corrects the leading monopole-monopole interaction between repeated
images, but leaves the monopole-dipole and dipole-dipole interactions
uncorrected. These two terms fall off like $1/V$ and $1/V^2$,  where $V$ is the cell size volume \cite{Makov_periodic_ewald_1995}.
To deal with this, we perform four  calculations at different
volumes, with the box size progressively increased by 1~\AA\
and fit the data to
\begin{equation}
\label{equ:boxconv}
 a_0 +a_1/V + a_2/V^2.
\end{equation}
For most molecules the corrections are small and only of the order of 10-20 meV,
whereas for the alkali dimers and some polar molecules the corrections can be as large as 100-200~meV.
In these cases the correction is very well described by the theoretical equation.
We hope to find a better solution in future work, for instance,
an explicit subtraction of monopole and dipole interactions between periodic images.
In terms of compute time, however, the additional calculations for smaller boxes
only require a modest amount of time:
since the total compute time scales quadratic to cubic with respect to the number of plane waves,
the calculations scale also quadratic to cubic in the volume. Typically we need 12~\AA\ large
boxes to obtain results converged to 20~meV with respect to the box size.
The additional smaller volumes used for the extrapolation require only half
of the compute time of the largest final box.

The final QP energies reported in the next section were obtained
by calculating the PBE one-electron HOMO and LUMO for a 25~\AA\ box
at an energy  cutoff that is 30~\% increased compared to the VASP default values.
The vacuum level, evaluated as the Hartree plus ionic potential, was evaluated
at the position furthest from the center of the molecule and subtracted from the
PBE one-electron energies.
We checked that the DFT one-electron energies are converged to
a few meV with this setup. To the DFT one-electron energies,
the shift of the QP energies $E^{\rm QP}-\epsilon^{\rm DFT}$
for the largest considered box, box size corrections, and basis set
corrections as described above are added.
It goes without saying that this procedure is rather involved and
since errors are expected to accumulate, we estimate that the present
predictions are only accurate to about $\pm50$~meV, where convergence
with cell size is the main source of errors and difficult to estimate precisely.

To give a feeling for the required compute time and computational effort,
we need to stress that our plane wave code is mainly designed for solids.
Nevertheless, a calculations for C$_6$H$_6$ in a 10~\AA\ box at the default cutoff
takes about 4 hours on a single node with 16 Xeon v2 cores.  The compute time stays roughly constant if the box size is increased by 1~\AA\ and the number
of cores is simultaneously doubled. Furthermore, the compute time is mostly independent of
the number of atoms in the box, but increases cubically with the box size
as the  total number of plane waves increases linearly with the box size.
By comparison Turbomole, using the def2-TZVP basis and the resolution of the identity method, takes 30 minutes on a 12 core 
 AMD opteron 6174 for the response and $GW$ part of the calculation 
(the time spent for the DFT part is negligible in comparison). 

\section{Results} \label{sec:results}

\subsection{HOMO for $GW$100}
\begin{table*}
\caption{
Ionization potential (IP, negative of HOMO QP energies) for 100 molecules using $G_0W_0$ and linearized  lin-$G_0W_0$ method. 
For comparison the basis set extrapolated values
of Ref. \onlinecite{Setten_GW100_2015} and the experimental IPs are given (vertical IPs are in italics).
If basis set extrapolated values are not specified in Ref.  \onlinecite{Setten_GW100_2015},
the AIMS-P16 values are shown in the column GTO (marked by $^*$).
Last column shows the  differences between GTO and PW results. The $^*$
indicates differences to non basis set extrapolated values.
}
\label{tab:HOMO}
\begin{ruledtabular}
\begin{tabular}{llrrrrr}
      &  &  $G_0W_0$ &   $G_0W_0$  &   lin-$G_0W_0$   & EXP & $\Delta$    \\
      &  & GTO\cite{Setten_GW100_2015} & PW & PW & & PW-GTO \\
\hline
  1 &              He &    23.49(0.03) &    23.38 &    23.62 & {\em 24.59} \cite{Kelly1987}  &     -0.11~ \\
  2 &              Ne &    20.33(0.01) &    20.17 &    20.36 & {\em 21.56} \cite{Kelly1987}  &     -0.16~ \\
  3 &              Ar &    15.28(0.03) &    15.32 &    15.42 & {\em 15.76} \cite{Weitzel1994} &     0.04~ \\
  4 &              Kr &    13.89(0.16) &    13.93 &    14.03 &    14.00 \cite{Wetzel1987}  &       0.04~ \\
  5 &              Xe &    12.02$^*$   &    12.14 &    12.22 &    12.13 \cite{Schafer1987} &    0.12$^*$ \\
  6 &    $ {\rm H_2}$ &    15.85(0.09) &    15.85 &    16.06 &    15.43 \cite{McCormack1989} &     0.00~ \\
  7 &   $ {\rm Li_2}$ &     5.05(0.02) &     5.09 &     5.32 &     4.73 \cite{Dugourd1992}   &     0.04~ \\ 
  8 &   $ {\rm Na_2}$ &     4.88(0.03) &     4.93 &     5.06 &     4.89 \cite{Kappes1985}    &     0.05~ \\ 
  9 &   $ {\rm Na_4}$ &     4.14(0.03) &     4.17 &     4.23 &     4.27 \cite{Herrmann1978}  &     0.03~ \\
 10 &   $ {\rm Na_6}$ &     4.34(0.06) &     4.34 &     4.40 &     4.12 \cite{Herrmann1978}  &     0.00~ \\
 11 &    $ {\rm K_2}$ &     4.08(0.04) &     4.12 &     4.24 &     4.06 \cite{Kappes1985}    &     0.04~ \\
 12 &   $ {\rm Rb_2}$ &     3.79$^*$   &     4.02 &     4.14 &     3.90 \cite{Kappes1985}  &    0.23$^*$ \\
 13 &    $ {\rm N_2}$ &    15.05(0.04) &    14.93 &    15.06 &    15.58 \cite{Trickl1989}    &    -0.12~ \\ 
 14 &    $ {\rm P_2}$ &    10.38(0.04) &    10.35 &    10.40 &    10.62 \cite{Bulgin1976}    &    -0.03~ \\ 
 15 &   $ {\rm As_2}$ &     9.67(0.10) &     9.59 &     9.62 &    10.0  \cite{Lau1982}       &    -0.08~ \\      
 16 &    $ {\rm F_2}$ &    15.10(0.04) &    14.93 &    15.08 &    15.70 \cite{Lonkhuyzen1984}&    -0.17~ \\ 
 17 &   $ {\rm Cl_2}$ &    11.31(0.05) &    11.32 &    11.40 & {\em 11.49} \cite{Dyke1984}    &     0.01~ \\ 
 18 &   $ {\rm Br_2}$ &    10.56(0.18) &    10.57 &    10.65 & {\em 10.51} \cite{Dyke1984}    &     0.01~ \\     
 19 &    $ {\rm I_2}$ &     9.23$^*$   &     9.52 &     9.59 & {\em 9.36} \cite{Kimura1981}   &     0.29$^*$ \\     
 20 &   $ {\rm CH_4}$ &    14.00(0.06) &    14.02 &    14.14 & {\em 13.6} \cite{Bieri1980}    &     0.02~ \\ 
 21 & $ {\rm C_2H_6}$ &    12.46(0.06) &    12.50 &    12.58 & {\em 11.99} \cite{Kimura1981}  &     0.04~ \\     
 22 & $ {\rm C_3H_8}$ &    11.89(0.06) &    11.90 &    11.98 & {\em 11.51} \cite{Kimura1981}  &     0.01~ \\     
 23 & ${\rm C_4H_{10}}$ &    11.59(0.05) &    11.61 &    11.69 & {\em 11.09} \cite{Kimura1981}  &   0.02~ \\     
 24 & $ {\rm C_2H_4}$ &    10.40(0.03) &    10.42 &    10.50 & {\em 10.68} \cite{Bieri1980}   &     0.02~ \\ 
 25 & $ {\rm C_2H_2}$ &    11.09(0.01) &    11.07 &    11.24 & {\em 11.49} \cite{Bieri1980}   &    -0.02~ \\ 
 26 &    $ {\rm C_4}$ &    10.91(0.03) &    10.89 &    10.97 &    12.54 \cite{Ramanathan1993} &   -0.02~ \\      
 27 & $ {\rm C_3H_6}$ &    10.65(0.04) &    10.72 &    10.78 & {\em 10.54} \cite{Plemenkov1981} &   0.07~ \\     
 28 & $ {\rm C_6H_6}$ &     9.10(0.01) &     9.11 &     9.16 & {\em 9.23} \cite{Howell1984}   &     0.01~ \\ 
 29 & $ {\rm C_8H_8}$ &     8.18(0.02) &     8.19 &     8.24 & {\em 8.43} \cite{Fu1978}    &        0.01~ \\
 30 & $ {\rm C_5H_6}$ &     8.45(0.02) &     8.47 &     8.51 & {\em 8.53} \cite{Kiselev1992}  &     0.02~ \\
 31 & ${\rm CH_2CHF}$ &    10.32(0.02) &    10.28 &    10.36 & {\em 10.63} \cite{Bieri1981}   &    -0.04~ \\
 32 & ${\rm CH_2CHCl}$ &     9.89(0.02) &     9.92 &    10.00 & {\em 10.20} \cite{Cambi1983}   &    0.03~ \\
 33 & ${\rm CH_2CHBr}$ &     9.14(0.01) &     9.75 &     9.83 & {\em 9.90} \cite{Cambi1983}    &    0.61~ \\
 34 & ${\rm CH_2CHI}$ &     9.01$^*$   &     9.27 &     9.36 & {\em 9.35} \cite{Wittel1974}  &   0.26$^*$ \\
 35 & ${\rm CF_4}$ &    15.60(0.06) &    15.41 &    15.53 & {\em 16.20} \cite{Bieri1981a} &    -0.19~ \\
 36 &  ${\rm CCl_4}$ &    11.21(0.06) &    11.20 &    11.31 & {\em 11.69} \cite{Kimura1981} &    -0.01~ \\
 37 &  ${\rm CBr_4}$ &    10.22(0.16) &    10.25 &    10.38 & {\em 10.54} \cite{Dixon1971}  &     0.03~ \\
 38 &  ${\rm CI_4}$ &     8.71$^*$   &     9.11 &     9.23 & {\em 9.10} \cite{Jonkers1982} &   0.40$^*$ \\
 39 &  ${\rm SiH_4}$ &    12.40(0.06) &    12.40 &    12.53 & {\em 12.3} \cite{Roberge1978} &     0.00~ \\  
 40 &  ${\rm GeH_4}$ &    12.11(0.04) &    12.13 &    12.24 & {\em 11.34} \cite{Potts1972}  &     0.02~ \\        
 41 &  ${\rm H_6Si_2}$ &   10.41(0.06) &    10.44 &    10.52 & {\em 10.53} \cite{Bock1976}   &     0.03~ \\ 
 42 & ${\rm H_{12}Si_5}$ &   9.05(0.05) &     9.13 &     9.19 & {\em 9.36} \cite{Bock1976}  &     0.08~ \\      
 43 &             LiH &     6.58(0.04) &     6.46 &     7.20 &     7.90 \cite{NIST2015}    &    -0.12~ \\         
 44 &              KH &     4.99(0.01) &     4.97 &     5.37 &     8.00 \cite{Farber1982}  &    -0.02~ \\         
 45 &    ${\rm BH_3}$ &    12.96(0.06) &    12.95 &    13.09 &    12.03 \cite{Ruscic1988}  &    -0.01~ \\         
 46 &  ${\rm B_2H_6}$ &    11.93(0.06) &    11.94 &    12.04 & {\em 11.90} \cite{Asbrink1981} &  0.01~ \\         
 47 &    ${\rm NH_3}$ &    10.39(0.05) &    10.32 &    10.44 & {\em 10.82} \cite{Baumgaertel1989} & -0.07~ \\ 
 48 &    ${\rm HN_3}$ &    10.55(0.02) &    10.50 &    10.56 & {\em 10.72} \cite{Cvitas1976}  &    -0.05~ \\      
 \end{tabular}
\end{ruledtabular}
\end{table*}
\begin{table*}
\begin{ruledtabular}
\begin{tabular}{llrrrrr}
      &  &  $G_0W_0$ &   $G_0W_0$  &   lin-$G_0W_0$   & EXP & $\Delta$    \\
      &  & GTO & PW & PW & & PW-GTO \\
\hline
 49 &    ${\rm PH_3}$ &    10.35(0.05) &    10.35 &    10.45 & {\em 10.59} \cite{Cowley1982}  &     0.00~ \\ 
 50 &   ${\rm AsH_3}$ &    10.21(0.02) &    10.26 &    10.36 & {\em 10.58} \cite{Demuth1977}  &     0.05~ \\
 51 &    ${\rm H_2S}$ &    10.13(0.04) &    10.11 &    10.30 & {\em 10.50} \cite{Bieri1982} &    -0.02~ \\  
 52 &              HF &    15.37(0.01) &    15.37 &    15.38 & {\em 16.12} \cite{Banna1975} &     0.00~ \\  
 53 &             HCl &    12.36(0.01) &    12.45 &    12.51 &    12.79 \cite{Wang1984}   &     0.09~ \\   
 54 &             LiF &    10.27(0.03) &    10.07 &    10.45 &    11.30 \cite{Berkowitz1962} &    -0.20~ \\
 55 &   ${\rm MgF_2}$ &    12.50(0.06) &    12.41 &    12.77 &    13.30 \cite{Hildenbrand1968} &  -0.09~ \\
 56 &   ${\rm TiF_4}$ &    14.07(0.05) &    14.01 &    14.22 & {\em 15.30} \cite{Dyke1997} &    -0.06~ \\
 57 &   ${\rm AlF_3}$ &    14.48(0.06) &    14.33 &    14.53 &    15.45 \cite{Dyke1984a} &    -0.15~ \\
 58 &              BF &    10.73(0.05) &    10.46 &    10.67 &    11.00 \cite{Farber1984} &    -0.27~ \\
 59 &    ${\rm SF_4}$ &    12.38(0.07) &    12.20 &    12.29 &    11.69 \cite{Fisher1992} &    -0.18~ \\
 60 &             KBr &     7.57(0.13) &     7.80 &     8.04 & {\em 8.82} \cite{Potts1977} &     0.23~ \\
 61 &            GaCl &     9.74(0.07) &     9.89 &     9.99 & {\em 10.07} \cite{Grabandt1990} & 0.15~ \\
 62 &            NaCl &     8.43(0.14) &     8.47 &     8.76 & {\em 9.80} \cite{Potts1977} &     0.04~ \\   
 63 &  ${\rm MgCl_2}$ &    11.20(0.07) &    11.19 &    11.41 & {\em 11.80} \cite{Lee1979} &    -0.01~ \\
 64 &   ${\rm AlI_3}$ &     9.30$^*$   &     9.58 &     9.69 & {\em 9.66} \cite{Barker1975} &   0.28$^*$ \\
 65 &              BN &    11.15(0.03) &        - &    10.61 &  11.50               &    - ~ \\
 66 &             HCN &    13.32(0.01) &    13.29 &    13.43 & {\em 13.61} \cite{Kreile1982} &    -0.03~ \\ 
 67 &              PN &    11.29(0.04) &    11.24 &    11.41 &    11.88 \cite{Bulgin1977} &    -0.05~ \\
 68 &  ${\rm N_2H_4}$ &     9.37(0.04) &     9.33 &     9.45 & {\em 8.98} \cite{Vovna1975} &    -0.04~ \\   
 69 &   ${\rm H_2CO}$ &    10.46(0.02) &    10.42 &    10.57 &    10.88 \cite{Ohno1995} &    -0.04~ \\     
 70 &  ${\rm CH_3OH}$ &    10.67(0.05) &    10.61 &    10.72 & {\em 10.96} \cite{Vorobev1989} &    -0.06~ \\ 
 71 & ${\rm CH_3CH_2OH}$ &    10.27(0.05) &    10.21 &    10.33 & {\em 10.64} \cite{Ohno1985} &    -0.06~ \\
 72 & ${\rm CH_3CHO}$ &     9.66(0.03) &     9.63 &     9.80 & {\em 10.24} \cite{Johnson1982} &    -0.03~ \\
 73 & ${\rm CH_3CH_2OCH_2CH_3}$ &  9.42(0.05) &     9.43 &     9.52 & {\em 9.61} \cite{Ohno1985} &  0.01~ \\
 74 &           HCOOH &    10.87(0.01) &    10.81 &    10.98 & {\em 11.50} \cite{VonNiessen1980} & -0.06~ \\
 75 &  ${\rm H_2O_2}$ &    11.10(0.01) &    10.96 &    11.12 & {\em 11.70} \cite{Ashmore1977} &    -0.14~ \\
 76 &    ${\rm H_2O}$ &    12.05(0.03) &    11.84 &    12.05 & {\em 12.62} \cite{Kimura1981} &    -0.21~ \\ 
 77 &    ${\rm CO_2}$ &    13.46(0.06) &    13.36 &    13.44 & {\em 13.77} \cite{Eland1977} &    -0.10~ \\  
 78 &    ${\rm CS_2}$ &     9.95(0.05) &     9.96 &    10.01 & {\em 10.09} \cite{Schweig1974} &     0.01~ \\
 79 &             CSO &    11.11(0.05) &    11.06 &    11.13 & {\em 11.19} \cite{Potts1974} &    -0.05~ \\
 80 &            COSe &    10.43(0.09) &    10.42 &    10.50 &    10.37 \cite{Cradock1975} &    -0.01~ \\
 81 &              CO &    13.71(0.04) &    13.62 &    13.76 & {\em 14.01} \cite{Potts1974} &    -0.09~ \\  
 82 &     ${\rm O_3}$ &    11.49(0.03) &    -     &    12.07 & {\em 12.73} \cite{Katsumata1984} &  - ~   \\
 83 &    ${\rm SO_2}$ &    12.06(0.06) &    11.91 &    12.04 & {\em 12.50} \cite{Kimura1981} &    -0.15~ \\ 
 84 &             BeO &     8.60(0.01) &     -    &     9.50 &    10.10 \cite{Theard1964} &     - ~   \\
 85 &             MgO &     6.75(0.03) &     -    &     7.10 &     8.76 \cite{Dalleska1994a} &     - ~   \\
 86 & ${\rm C_6H_5CH_3}$ &     8.73(0.02) &     8.75 &     8.79 & {\em 8.82} \cite{Howell1984} &   0.02~ \\
 87 &  ${\rm C_8H_{10}}$ &     8.66(0.02) &     8.69 &     8.73 & {\em 8.77} \cite{Howell1984} &   0.03~ \\
 88 &     ${\rm C_6F_6}$ &     9.74(0.07) &     9.63 &     9.69 & {\em 10.20} \cite{Bieri1981a} &  -0.11~ \\
 89 &     ${\rm C_6H_5OH}$ &     8.51(0.01) &     8.38 &     8.43 &     8.75 \cite{Ballard1987a} & -0.13~ \\
 90 &   ${\rm C_6H_5NH_2}$ &     7.78(0.01) &     7.78 &     7.84 & {\em 8.05} \cite{Furin1987} &   0.00~ \\
 91 &   ${\rm C_5H_5N}$ &     9.17(0.01) &     9.16 &     9.31 & {\em 9.66} \cite{Kobayashi1974} & -0.01~ \\
 92 &         Guanine &     7.87(0.01) &     7.85 &     7.90 & {\em 8.24} \cite{Hush1975} &    -0.02~ \\
 93 &         Adenine &     8.16(0.01) &     8.12 &     8.18 & {\em 8.48} \cite{Lin1980} &    -0.04~ \\
 94 &        Cytosine &     8.44(0.01) &     8.40 &     8.50 & {\em 8.94} \cite{Hush1975} &    -0.04~ \\
 95 &         Thymine &     8.87(0.01) &     8.83 &     8.89 & {\em 9.20} \cite{Dougherty1976} &  -0.04~ \\
 96 &          Uracil &     9.38(0.01) &     9.36 &     9.55 & {\em 9.68} \cite{Palmer1980} &    -0.02~ \\
 97 & ${\rm NH_2CONH_2}$ &     9.46(0.02) &     9.35 &     9.59 & {\em 9.80} \cite{Bieri1982} &    -0.11~ \\
 98 &     ${\rm Ag_2}$ &     7.08$^*$   &     7.83 &     7.95 &     7.66 \cite{Beutel1993} &     0.75$^*$ \\
 99 &     ${\rm Cu_2}$ &     7.78(0.06) &     7.19 &     7.40 &     7.46 \cite{Franzreb1990} &    -0.59~ \\
100 &            CuCN &     9.56(0.04) &     -    &     9.99 &     -   &     -   ~ \\
\end{tabular}
\end{ruledtabular}
\end{table*}

Let us first note on the agreement at the level of DFT (not shown).
In general, our values agree exceedingly well with the PBE values
reported in the supplementary material of Ref. \onlinecite{Setten_GW100_2015}. In most
cases, our PBE HOMO is located between the basis set extrapolated
values and the values obtained with the best basis sets used in
the GTO calculations (def2-QZVP). On average, our PW HOMOs agree better
with the non basis set extrapolated values with a mean deviation (MD) of 7~meV
and a mean absolute deviation (MAD) of 19~meV. Compared to the GTO basis set
extrapolated values, the MD and MAD are -25~meV and 30~meV 
(in both cases, CH$_2$CHBr was excluded, see below).

van Setten {\em et al.} \cite{Setten_GW100_2015} extrapolated
the DFT eigenvalues using a cubic polynomial in the inverse of the basis set cardinal number ($C_n^{-3}$);
we believe  that this is not appropriate and will overestimate the basis set corrections.
It is commonly agreed that DFT calculations converge exponentially
with the  cardinal number, whereas any correlated wave function
calculation converges with the inverse of the basis set size (corresponding
roughly to $C_n^{-3}$) \cite{Bruneval2013,Klimes_predictiveGW_2014,Setten_GW100_2015}.
This is a result of Kato's cusp condition \cite{Kato_CUSP_57}
causing a kink in the many-body wave function as two coordinates
approach each other. We have shown that this problem carries over
to $GW$ calculations \cite{Klimes_predictiveGW_2014}.
As a one-electron theory, density functional theory does not suffer
from this slow convergence.
We hence believe that van Setten overestimated
the basis set corrections for DFT. This is supported by the observation that our PBE results 
tend to be closer to the non-extrapolated Gaussian results at the level of def2-QZVP.

 We now turn to the QP energies predicted at the level of $G_0W_0$ shown in Table \ref{tab:HOMO}.
The  agreement between the VASP PW and the GTO results is generally very good.
We note that the $G_0W_0$ approximation used here is identical
to the one applied by van Setten {\em et al.} \cite{Setten_GW100_2015}.
Specifically, van Setten determined the nodes  of Eq. (\ref{equ:QPfull}),
and we do exactly the same in the present work. Linearization of the
QP equation (\ref{equ:QPlinear}) yields generally somewhat larger QP energies and
often improves agreement with experiment slightly (column  lin-$G_0W_0$).
This trend has also been observed in a recent benchmark for an unrelated set of molecules \cite{Scherpelz2016}. 
In agreement with van Setten \cite{Setten_GW100_2015}, we have found poles in the self-energy close to the predicted
QP energies for BN, O$_3$, BeO, MgO and CuCN. Since analytic continuation
has difficulties to resolve the precise pole structure of the Green's function, we only report the values
obtained from the linearized self-energy.

Our discussion starts with the molecules that show large discrepancies
between VASP and GTO's.
A large out-liner is seemingly  CH$_2$CHBr. However,  for this molecule, as well as
C$_6$H$_5$OH, we found large forces in the preparatory PBE calculations.
Double checking the original literature~\cite{C2H4Br_str2011} suggests that the $GW$100 paper used
incorrect geometries. Since the ultimate purpose is certainly to compare with experiment,
we decided to update the geometries to the correct literature values.

\begin{table}
\caption{
IP (negative HOMO $G_0W_0$ QP energies) and $G_0W_0$ LUMO for selected molecules. 
The GTO values have been calculated using frozen core potentials, scalar relativistic corrections,
and are extrapolated to the infinite basis set limit. 
}
\label{tab:QPupdates}
\begin{ruledtabular}
\begin{tabular}{llrrrr}
      &   &  IP      &  IP   & LUMO & LUMO \\
      &  & GTO & PW   & GTO & PW \\
\hline
 5  &     Xe             &       12.22 &    12.14  & -0.07  &  0.28 \\
 12  &     ${\rm Rb_2}$  &       4.07  &     4.02  & -0.85  & -0.74 \\
 19 &     ${\rm I_2}$    &        9.48 &     9.52  & -2.28  & -2.21 \\
 34 &    ${\rm CH_2CHI}$ &        9.13 &     9.27  &  0.56  &  0.37 \\
 38 &     ${\rm CI_4}$    &       8.97 &     9.11  & -2.47  & -2.42 \\
 64 &       ${\rm AlI_3}$ &       9.50 &     9.58  & -1.18  & -1.02 \\
 98 &    ${\rm Ag_2}$     &       7.96 &     7.83  & -1.40  & -1.35  
\end{tabular}
\end{ruledtabular}
\end{table}

Among the remaining molecules, errors are large for compounds containing iodine, rubidium, and silver with a maximum
deviation of 400 meV for CI$_4$ and Rb$_2$, and 750~meV for Ag$_2$.
However, in Ref. \onlinecite{Setten_GW100_2015} no basis set extrapolation was performed for these molecules.
From CCl$_4$ to CBr$_4$, the
basis set corrections increase from 300 meV to 350 meV, suggesting
a basis set  error of 400 meV for CI$_4$ using GTOs. 
Similarly, for Rb$_2$ the GTO results were not basis set corrected,
and estimating the basis set error from Na$_2$ and K$_2$
again suggests  that the VASP results are accurate.
For Ag$_2$, the difference between VASP and GTO seem on first sight to be too
large to be ascribed to basis set errors alone. To resolve the issue, one of us (MvS) repeated the 
Xe, Rb$_2$, I$_2$, ${\rm CH_2CHI}$, CI$_4$, AlI$_3$, and Ag$_2$  calculations using scalar relativistic corrections and 
frozen core SVP, TZVP and QZVP basis sets. This yielded  basis set extrapolated
values summarized in Tab. \ref{tab:QPupdates} certainly now in good to very good agreement with
the VASP values.

\begin{table}
\caption{
IP (negative HOMO $G_0W_0$ QP energies) for selected molecules calculated for a 9~\AA\ box for
the standard $GW$ potentials and normconserving $GW$ potentials. Results differ
from the previous table, since calculations in Table \ref{tab:HOMO} have
been performed for larger boxes and include a correction for the box size error.
The column $\Delta$ reports the difference between the standard PAW and NC PAW potential.
}
\label{tab:QPNC}
\begin{ruledtabular}
\begin{tabular}{llrrrrr}
    &                 & $GW$ PAW     & NC $GW$ PAW & $\Delta$    \\
\hline
 13 &     ${\rm N_2}$ &       14.98 &    15.02 &        -0.04 \\
 16 &     ${\rm F_2}$ &       14.97 &    15.13 &        -0.15 \\
 35 &    ${\rm CF_4}$ &       15.42 &    15.58 &        -0.16 \\
 52 &              HF &       15.39 &    15.32 &         0.06 \\
 58 &              BF &       10.42 &    10.46 &        -0.04 \\
 59 &    ${\rm SF_4}$ &       12.19 &    12.26 &        -0.07 \\
 76 &    ${\rm H_2O}$ &       11.86 &    11.94 &        -0.09 \\
 99 &    ${\rm Cu_2}$ &        7.03 &     7.53 &        -0.50 \\
\end{tabular}
\end{ruledtabular}
\end{table}

 For the remaining molecules, the mean absolute deviation between the
two codes and thus two completely different basis sets is only 60 meV, if we also exclude Cu$_2$.
For Cu$_2$, the fluorine containing compounds, H$_2$O, as well as Ne
the ionization potentials (IPs) are smaller in VASP, which we will now show to be related
to slight deficiencies in the PAW potentials. Copper, neon,
and fluorine and, to a lesser extent, oxygen are particularly difficult to describe using a plane wave based approach,
since the $3d$ and $2p$ electrons are strongly localized. To cope with this, the Cu, F
and Ne potentials are already the three smallest core and hardest potentials used in the present
work. But still, the partial waves do not conserve the norm exactly, which
results in errors, if an electron is scattered into a plane wave with very high kinetic energy \cite{Klimes_predictiveGW_2014}.
To determine this error, we performed calculations with norm-conserving (or almost norm-conserving)
 $GW$ potentials for the molecules Cu$_2$, N$_2$, F$_2$, CF$_4$, HF, BF,  SF$_4$ and H$_2$O reported in Tab. \ref{tab:QPNC} 
using the potentials {\tt Cu$\_$sv$\_$GW$\_$nc}, {\tt N$\_$h$\_$GW},...,{\tt F$\_$h$\_$GW}. 
Except for HF,
the QP energies are clearly shifted towards higher binding energies in these calculations, and
the discrepancies to the GTO calculations are reduced to an acceptable level of 100 meV. We also note that
the PAW error  increases from nitrogen, over oxygen to fluorine. HF and BF are
exceptions, since the  HOMOs possess predominantly hydrogen and boron character and, therefore, do
not depend strongly on the F potential (we note that the HF results were already accurate using
the standard potentials). Finally, the standard carbon and boron potentials
used here are already almost norm conserving, and hence negligible changes are found
for carbon based compounds with harder potentials (not shown).

The final case worthwhile mentioning is KBr. Here the $GW$100 paper \cite{Setten_GW100_2015}
reports relatively large extrapolation errors of 130~meV, indicating that
in this case the GTO based extrapolation might be inaccurate.

For the remaining systems, we find the agreement to be excellent. Specifically, for all considered
organic molecules the absolute differences  are typically below 50 meV, with very few
out-liners. This clearly demonstrates that plane wave codes can be competitive
in terms of precision with GTOs. Certainly the agreement between GTOs and PWs
is better than originally reported in the $GW$100 paper, a point discussed
in more detail in the next section.

\begin{figure}
    \begin{center}
      ~  \includegraphics[width=8cm,clip=true]{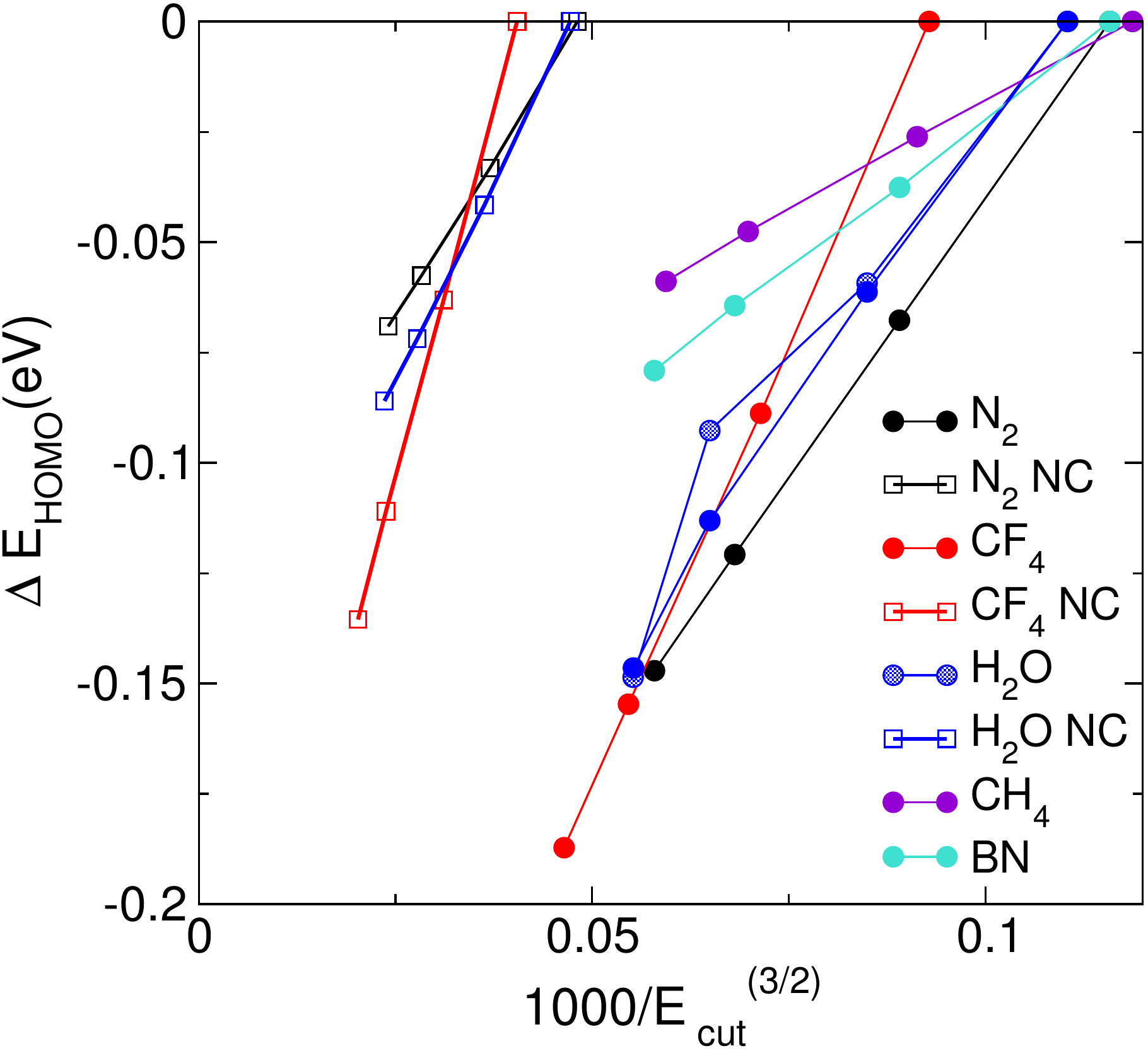}
    \end{center}
   \caption{
\label{fig:cutoff}
(color online)
Convergence of QP HOMO with respect to the employed cutoff
for various materials. For N$_2$, CF$_4$ and H$_2$O, results
are shown for two potentials, the standard $GW$ potentials, as
well as NC potentials. The slopes are steeper for
the NC potentials, which is particularly obvious for CF$_4$.
For H$_2$O results for two box sizes (8 and 9~\AA) are shown (see text).
}
\label{fig:goldstone}
\end{figure}
\subsection{Basis set convergence and comparison to other PW calculations}
\label{sec:convergencePW}
In Fig. \ref{fig:cutoff}, we show the convergence of the HOMO
with respect to the plane wave cutoff for the orbitals. This cutoff also
determines the total number of orbitals  as well as the
cutoff for the response function. The number of plane waves and total orbitals
is proportional to one over the cutoff to the power of 3/2. It is clearly
visible that the curves follow almost exactly a straight line.
In few cases, out-liners are visible. For instance for H$_2$O,
we have included results for two box sizes 8 and 9~\AA.
The 9~\AA\ box results have a slight jump, which is not present for
the 8~\AA\ box. However, this out-liner is small (about 10 meV), and
changes extrapolated results only by less than 10 meV. Usually
the out-liners could be dealt with by just changing the box size
slightly. We believe that they are related to "shell"-effects, {\em i.e.}
a sudden increase in the number of plane waves when the cutoff is changed
through certain values. Furthermore,  the analytic continuation is
not always entirely well behaved and can cause changes of the order
of 20 meV. Overall, the plot  demonstrates
that extrapolation with respect to the energy-cutoff is very
well under control and can be done with great accuracy.

As noted before, the differences between the GTO and  the Berkeley $GW$ calculations reported in Ref. \onlinecite{Setten_GW100_2015}
are more sizable. If we exclude the difficult multipole cases, BN, O$_3$, MgO and BeO,
the mean absolute deviation between Berkeley $GW$ and basis set extrapolated GTOs
was 0.2 eV in Ref.  \onlinecite{Setten_GW100_2015},
whereas it is reduced to 0.05 eV for VASP PAW potentials (for the same subset).
We speculate that this is mostly related to neglecting basis set
extrapolation errors or--- less likely ---to an inaccurate treatment of
the core-valence interaction. Typically our basis set corrections are of the order of 300-400 meV
at the default cutoff and therefore very sizable. Even doubling the number of basis functions
and therefore increasing the compute time by a factor of about 8 (cubic scaling),
reduces the error only by a factor 2, to about 150-200 meV. Hence,
calculations without basis set corrections are hardly affordable or practicable, and it is
certainly advisable to perform an extrapolation whenever possible.

\begin{table*}
\caption{
LUMO QP energies for selected molecules using $G_0W_0$ and the linearized  lin-$G_0W_0$. For comparison, the
non basis set extrapolated values (AIMS-P16), the basis set extrapolated values
from Ref. \onlinecite{Setten_GW100_2015}, and
the negative of the experimental electron affinities are shown (vertical attachment energies are in italics). Differences between PW and GTO
are shown in last column. The $^+$ indicates energies above the vacuum level, and
$^*$ indicates differences to non basis set extrapolated values. }
\label{tab:LUMO}
\begin{ruledtabular}
\begin{tabular}{llrrrrrr}
      &  &  $G_0W_0$ &  $G_0W_0$ &   $G_0W_0$  &   lin-$G_0W_0$   & EXP & $\Delta$    \\
      &  &    AIMS-P16                   & GTO-EXTRA & PW & PW & & PW-GTO \\
\hline
  2 &             Ne &    11.64 &      -       &     0.40   &    0.40 &     - &   -11.24$^+$ \\
  5 &             Xe &     4.28 &        -     &     0.70   &    0.70 &     - &    -3.58$^+$ \\ 
  6 &    ${\rm H_2}$ &     3.50 &   3.30(0.52) &     0.07   &    0.07 &     - &    -3.23$^+$ \\
  7 &   ${\rm Li_2}$ &    -0.63 &  -0.75(0.04) &    -0.61   &   -0.54 &     - &     0.14~ \\
  8 &   ${\rm Na_2}$ &    -0.55 &  -0.66(0.70) &    -0.60   &   -0.56 &     -0.54 & 0.06~ \\
  9 &   ${\rm Na_4}$ &    -1.01 &  -1.15(0.90) &    -1.07   &   -1.03 &     -0.91 \cite{McHugh1989} & 0.08~ \\
 10 &   ${\rm Na_6}$ &    -0.97 &  -1.13(0.10) &    -1.07   &   -1.03 &     - &     0.06~ \\
 11 &    ${\rm K_2}$ &    -0.65 &  -0.75(0.05) &    -0.74   &   -0.70 &     -0.50 &     0.01~ \\
 12 &   ${\rm Rb_2}$ &    -0.62 &        -     &    -0.74   &   -0.70 &     -0.50 \cite{McHugh1989} & -0.12$^*$ \\
 14 &    ${\rm P_2}$ &    -0.72 &  -1.08(0.08) &    -0.99   &   -0.97 &     -0.68 \cite{Jones1995} &  0.09~ \\
 15 &   ${\rm As_2}$ &    -0.85 &  -1.52(0.35) &    -1.07   &   -1.06 &     -0.74 \cite{Lippa1998} &  0.45~ \\
 16 &    ${\rm F_2}$ &    -0.70 &  -1.23(0.14) &    -0.96   &   -0.84 &    {\em -1.24} \cite{Ayala1981} &  0.27~ \\
 17 &   ${\rm Cl_2}$ &    -0.89 &  -1.40(0.12) &    -1.25   &   -1.22 &    {\em -1.02} \cite{Ayala1981} &  0.15~ \\
 18 &   ${\rm Br_2}$ &    -1.40 &  -1.96(0.29) &    -1.99   &   -1.97 &    {\em -1.60} \cite{Ayala1981} & -0.03~ \\
 19 &    ${\rm I_2}$ &    -1.68 &        -     &    -2.21   &   -2.20 &    {\em -1.70} \cite{Ayala1981} & -0.53$^*$ \\
 20 &   ${\rm CH_4}$ &     2.45 &   2.03(0.35) &     0.63   &    0.63 &     - &    -1.40$^+$ \\
 26 & ${\rm C_4}$    &    -2.94 &  -3.15(0.06) &    -3.09   &   -3.08 &     -3.88 \cite{Arnold1991} &   0.06~ \\
 29 & ${\rm C_8H_8}$ &     0.06 &  -0.12(0.02) &    -0.05   &   -0.02 &     -0.57 \cite{Miller2002} &     0.07~ \\
 32 & ${\rm CH_2CHCl}$ &   1.42 &   1.17(0.03) &     1.19   &    1.25 &     - &     0.02$^+$ \\
 36 & ${\rm CCl_4}$ &     -0.01 &  -0.54(0.13) &    -0.32   &   -0.28 &     - &     0.22~ \\
 37 & ${\rm CBr_4}$ &     -1.08 &  -1.56(0.29) &    -1.47   &   -1.44 &     - &     0.09~ \\
 38 & ${\rm CI_4}$ &      -2.14 &        -     &    -2.42   &   -2.40 &     - &    -0.28$^*$ \\
 42 & ${\rm H_{12}Si_5}$ & 0.16 &   0.00(0.07) &     0.03   &    0.05 &     - &     0.03~ \\
 43 &       LiH      &    -0.07 &  -0.16(0.09) &    -0.07   &   -0.04 &     -0.34 \cite{Sarkas1994} &     0.09~ \\
 44 &            KH  &    -0.18 &  -0.32(0.01) &    -0.25   &   -0.22 &     - &     0.07~ \\
 45 &  ${\rm BH_3}$  &     0.12 &   0.03(0.05) &     0.03   &    0.08 &     -0.04 \cite{WickhamJones1989} &     0.00~ \\
 54 &           LiF  &     0.09 &  -0.01(0.01) &     0.17   &    0.17 &     - &     0.18~ \\
 55 &  ${\rm MgF_2}$ &    -0.14 &  -0.31(0.06) &    -0.29   &   -0.28 &     - &     0.02~ \\
 56 &  ${\rm TiF_4}$ &    -0.60 &  -1.06(0.13) &    -0.79   &   -0.66 &     -2.50 \cite{Boltalina1991} &     0.27~ \\
 57 &  ${\rm AlF_3}$ &     0.16 &  -0.23(0.10) &     0.08   &    0.09 &     - &     0.31~ \\
 59 &   ${\rm SF_4}$ &     0.38 &  -0.10(0.13) &     0.07   &    0.12 &     -1.50 \cite{Miller1995} &     0.17~ \\
 60 &            KBr &    -0.31 &  -0.42(0.06) &    -0.32   &   -0.31 &     -0.64 \cite{Miller1986} &     0.10~ \\
 61 &           GaCl &    -0.02 &  -0.39(0.15) &    -0.19   &   -0.15 &     - &     0.20~ \\
 62 &           NaCl &    -0.39 &  -0.42(0.01) &    -0.46   &   -0.43 &     -0.73 \cite{Miller1986} &    -0.04~ \\
 63 & ${\rm MgCl_2}$ &    -0.43 &  -0.68(0.08) &    -0.61   &   -0.59 &     - &     0.07~ \\
 64 &  ${\rm AlI_3}$ &    -0.80 &        -     &    -1.02   &   -0.99 &     - &    -0.22$^*$ \\
 72 & ${\rm CH_3CHO}$ &    1.05 &   0.83(0.05) &     0.87   &    0.87 &     - &     0.04$^+$ \\
 74 &          HCOOH &     1.91 &   1.59(0.00) &     1.64   &    1.72 &     - &     0.05$^+$ \\
 76 &   ${\rm H_2O}$ &     2.37 &   2.01(0.16) &     1.04   &    1.04 &     - &    -0.97$^+$ \\
 78 &   ${\rm CS_2}$ &    -0.20 &  -0.55(0.09) &    -0.42   &   -0.40 &     -0.55 \cite{Cavanagh2012} &     0.13~ \\
 82 &    ${\rm O_3}$ &    -2.30 &  -2.69(0.11) &    -2.50   &   -2.52 &     -2.10 \cite{Arnold1994} &     0.19~ \\
 83 &   ${\rm SO_2}$ &    -1.00 &  -1.49(0.12) &    -1.25   &   -1.19 &     -1.11 \cite{Nimlos1986} &     0.24~ \\
 84 &            BeO &    -2.56 &  -2.72(0.04) &    -2.73   &   -2.37 &     - &    -0.01~ \\
 85 &            MgO &    -1.89 &  -2.13(0.09) &    -2.05   &   -2.12 &     - &     0.08~ \\
 88 & ${\rm C_6F_6}$ &     0.66 &   0.36(0.08) &     0.24   &    0.27 &     -0.70 \cite{Eustis2007} &    -0.12$^+$ \\
 94 &       Cytosine &     0.26 &   0.01(0.01) &     0.12   &    0.15 &        -0.23 \cite{Schiedt1998} &     0.11~ \\
 95 &        Thymine &     0.06 &  -0.18(0.01) &    -0.06   &   -0.04 & {\em 0.29} \cite{Aflatooni1998} &     0.12~ \\
 96 &         Uracil &     0.01 &  -0.25(0.01) &    -0.11   &   -0.09 & {\em 0.22} \cite{Aflatooni1998} &     0.14~ \\
 98 &   ${\rm Ag_2}$ &    -1.05 &      -       &    -1.35   &   -1.31 & {\em -1.10} \cite{Handschuh1995} &    -0.30$^*$ \\
 99 &   ${\rm Cu_2}$ &    -0.92 &  -1.23(0.08) &    -1.24   &   -1.21 &     -0.84 \cite{Taylor1992} &    -0.01~ \\
100 &           CuCN &    -1.65 &  -1.85(0.05) &    -1.91   &   -1.81 & {\em -1.47} \cite{Wu2010} &    -0.06~ \\
\end{tabular}
\end{ruledtabular}
\end{table*}

For the core-valence interaction, we emphasize that VASP always evaluates the interaction
at the level of Hartree-Fock if correlated
calculations are performed. More precisely, VASP calculates the PBE core orbitals
on the fly and then recalculates the action of the PBE core states on the valence
states using the Hartree-Fock approximation. Not doing so can have a sizable effect on the QP energies
for heavier atoms \cite{KresseGWa}. We are not aware of other pseudopotential codes following a similar route.
This might be responsible for a small part of the errors in the reported
Berkeley $GW$ calculations of Ref. \onlinecite{Setten_GW100_2015},
if heavier atoms are involved.

Calculations for another fairly large set of molecules have been reported by Govoni and Galli using
the West code\cite{Govoni_West_2015}.
29 molecules are identical to the $GW$100  set considered here.
The mean absolute difference between the basis set extrapolated GTO results
and the VASP results for this subset is 60~meV, whereas the difference between
the West results and the basis set extrapolated GTO results is about twice as large
120~meV (mean absolute difference between VASP and West is 90~meV).
In many cases, the West IPs are too small indicating again basis set
incompleteness errors. Anyhow, the West results are closer
to the basis set converged values than the $GW$ Berkeley results.

\subsection{Comparison to experiment}

When comparing the present results against the experimental ionisation energies, a mean absolute error of 0.5 eV is observed.
This large discrepancy is not unexpected given that in this computational approach self-consistency, vertex corrections and finite temperature effects are omitted.
However we can comment on the biggest outliers in the set. 
A first example is $\text{C}_\text{4}$: it is well known that the smaller $\text{C}_\text{2}$ molecule is particularly challenging to describe, owing to strong electron correlation \cite{Curtiss2007}.
For the larger cluster we expect similar effects, hence the inclusion of the vertex should improve the agreement with the experiment.
We have a similar expectation for the case of $\text{F}_\text{2}$.
Our conjecture is substantiated by previous electron propagator calculations \cite{Golab1986}, where the poles of the Green's function in the Lehmann representation were located to give the IP, and where a comparable mismatch to experiment was ascribed to the poor description of dynamic correlation.
For $\text{AlF}_\text{3}$, LiF and KH we have to bear in mind that the experimental value for a vertical transition was not available, therefore geometry relaxations may explain the mismatch.
This is only partially true for KH, where the inclusion of adiabatic effects in the perturbative calculations still leaves a sizable disagreement ($\sim$2 eV) \cite{Rayne2011}; in this case it is not completely unreasonable to call for a further assessment of the experimental value.

\subsection{Linearized QP-HOMO for $GW$100}\label{sec:linearized}

We now turn to results obtained by first linearizing the self-energy and
then determining the QP energy from this linearized equation. This procedure is
in our experience more "robust" and better behaved than seeking the
poles in the non-linearized equation. The main issue of the latter approach is that,
in the $G_0W_0$ approximation, the first pole in the self-energy is approximately located
at the energy of the  DFT HOMO minus the first excitation energy in the DFT (LUMO$-$HOMO):
\[
 \epsilon_{\rm HOMO}- ( \epsilon_{\rm LUMO}-\epsilon_{\rm HOMO}).
\]
This is a simple Auger like excitation, where the hole  has sufficient energy, {\em i.e.}
is sufficiently below the HOMO to be able to excite an electron-hole pair.
As discussed by van Setten, such poles lead to multiple solutions
for the QP energy \cite{Setten_GW100_2015} and make the determination of the QP energies
difficult for molecules with small excitation energies. These poles are, however, an artifact of the $G_0W_0$
approximation. If the $GW$ procedure were done self-consistently,
the first pole in the self-energy would move to approximately
\[
 E^{\rm QP}_{\rm HOMO}- ( E^{\rm QP}_{\rm LUMO}-E^{\rm QP}_{\rm HOMO}).
\]
In other words, at the valence band edge (HOMO) and conduction band edge (LUMO)
the self-energy never possesses poles. However, in a single shot procedure and when
starting from much too small band gaps,
the quasiparticle energy  $E^{\rm QP}$ might move into regions where the self-energy
evaluated from DFT orbitals has a pole. 
Linearization at the DFT eigenenergies
resolves this issue, as the $G_0W_0$ self-energy has no poles in the direct vicinity
of the DFT HOMO.  The problem is also less severe, if the calculations are
done selfconsistently or when starting from a prescription that yields larger HOMO-LUMO Kohn-Sham gaps, as shown
in a  recent evaluation  of the difference between the quasi-particle orbital 
energies and their linearized counterparts by Govoni {\em et al.} \cite{Govoni_West_2015}.
Therein it is shown that, for a wide range of molecules, this difference is substantially more pronounced for 
$GW$ calculations on a PBE reference state than if a  hybrid functional with non-local exchange is used.

In summary, we feel that for code benchmarking as well as for a comparison with experiment determining
the poles of the linearized equation is preferable, at least, if a PBE reference state is employed.
However, it also needs to be emphasized that for comparison with the 
already published $GW$100 data, it is of paramount importance to 
follow exactly the procedures laid out in the the initial $GW$100 paper.

\subsection{LUMO for $GW$100}

The calculated LUMOs are shown in Table \ref{tab:LUMO}. 
A few important comments are in place here. First, the table reports the QP energy
of the lowest unoccupied orbital in the preceeding DFT calculations to maintain compatibility
with the previous publication. In some cases (Xe, H$_2$O, CH$_2$CHCl, CH$_3$CHO, and HCOOH) PW calculations
 predict at the DFT level a very weakly bound LUMO+1 state (just below the vacuum level) whose $G_0W_0$ QP energy 
 is below the QP state corresponding to the DFT LUMO level. These energy levels are not shown in Tab. \ref{tab:LUMO}.

If we consider the $G_0W_0$ values corresponding to the DFT LUMOs, the agreement
between the $GW$100 reference GTO data and plane waves
is reasonable, although not quite as good as for the HOMO. Specifically troublesome is 
the observation that the GTO calculations sometimes predict much too  positive LUMOs
 several eV above the PW results. Admittedly, box size
convergence can be troublesome for QP energies above
the vacuum level, and we therefore only show few selected
positive LUMOs--- those where we are confident that convergence to
50 meV was attained for the cell sizes considered in our calculations. 
All positive (unbound) $G_0W_0$ LUMOs are marked by a superscript $"+"$ sign in the last column. 
The differences are particularly striking for Ne, Xe, H$_2$, H$_2$O and CH$_4$ reaching 11 eV for Ne. 

To investigate this issue, we compared the DFT-LUMOs of the PW and GTO 
calculations (the latter are available upon request to MvS), and found that the 
deviations between PWs and GTOs are much
larger than for the DFT HOMOs on average, and especially large for some of the 
problematic cases, e.g. Ne, Xe, H$_2$, or H$_2$O which are the largest outliners
in the subsequent QP calculations. Specifically, for Ne and H$_2$ the PW
DFT calculations predict very shallow bound LUMOs, a few 10 meV below the vacuum level.
These  can not be reproduced with any of the available  GTO basis sets. 

For the other cases with larger discrepencies, we now show that the GTO basis sets are often not 
sufficiently flexible to describe unoccupied orbitals. This is supported by several observations. 
(i)  Basis set corrections using GTOs are much larger for the LUMO than for the
HOMO, as for instance exemplified for As$_2$, F$_2$
or Cl$_2$. To make this very clear, we have included in Table \ref{tab:LUMO}
both, the basis set extrapolated values (with estimated error bars), as well as the values
at the largest considered GTO basis set. 
(ii) Non basis set extrapolated GTO values deviate markedly from PW results. 
As before, these are marked by a star superscript in the last column.  GTO basis set extrapolated values are tabulated in Tab. \ref{tab:QPupdates}
and clearly improve the agreement with the PW results. For Xe, where the discrepancy 
was previously 3.6~eV, the error is reduced to about 0.3~eV. 
Furthermore, we recalculated the QP energies of H$_2$O and CH$_4$ using 
Dunning correlation consistent basis sets and found basis set extrapolated $G_0W_0$ QP energies of 
1.00~eV  and 0.89~eV, now in excellent and reasonable agreement with the PW results. All in all,
we therefore conclude that the Gaussian basis set results for unoccupied states
need to be considered with some caution, and Dunning correlation consistent basis sets
are seemingly better suited to predict accurate values. 

If we restrict the comparison between PWs and GTOs to states below the vacuum level, 
we find the agreement to be generally, as for the HOMO, rather satisfactory. Differences are
about a factor two larger than for the HUMO, but considering the previous discussion
on the possible issues with the Gaussian basis sets for unoccupied orbitals, this is certainly
not astonishing.

Finally, concerning the agreement with the experiment, we find a similar absolute deviation as for the first 
ionization energies (compare with Tab. \ref{tab:HOMO}). To make the comparison between the LUMO energies 
and the experiment more immediate, the second last  column reports the negative of the experimental electron affinities, which is overall
in quite satisfactory agreement with experiment.

\section{Discussions and Conclusions}

The main purpose of the present work is a careful comparison of
 $GW$ QP energies obtained using Gaussian type orbitals and plane waves.
One important motivation was that the values reported for the Berkeley $GW$ code
were typically 200 meV smaller than the  basis set extrapolated
GTO results. However, the Gaussian basis set extrapolation also often increased
the predicted QP energies by some 100 meV. Since basis set extrapolation
using GTOs is not necessarily accurate,  and since the Berkeley $GW$
calculations are often closer to the uncorrected values
than the basis set extrapolated values, 
we felt that it is important to bring in a third independent set of calculations, hopefully confirming
one or the other of the previous values.

The main outcome of our work is that our VASP predicted HOMOs are in excellent agreement with the
basis set extrapolated GTO results. We believe this establishes beyond
doubt that the values reported in Ref. \onlinecite{Setten_GW100_2015}
are very reliable and can be used as a rigorous benchmark for future implementations.
In the few cases (iodine compounds, Br$_2$ and Ag$_2$), where the GTO calculations
were not extrapolated to the basis set limit, we find--- not unexpectedly ---that
the non basis set extrapolated GTO values underestimate the IP by about 300-400 meV.
The present work also reports basis set extrapolated GTO values for these molecules
finding good agreement with VASP PW results.

Although the mean absolute deviation between our PAW PW results and the GTO
results is only 60 meV, we found larger discrepancies for molecules containing copper,
fluorine  and nitrogen. We traced these differences back to the use
of non-normconserving PAW potentials: using normconserving PAW potentials the agreement
between PW calculations and GTOs improves further.

For the LUMO, results are slightly less satisfactory. Agreement between GTOs and PWs is 
good for QP energies below the vacuum level, although even for those there
are more out-liners and the average deviation is larger. For instance, differences are sizable for some
seemingly simple dimers. We attribute this to very large basis
set corrections  for GTOs  for some molecules  (e.g. 700 meV for As$_2$).

If the predicted QP LUMOs are above the vacuum level, the differences
between the PW and GTO results can be very large and can
reach  11~eV (Ne). 
Our explanation for this behavior is that the GTO basis sets employed in Ref. \onlinecite{Setten_GW100_2015}
are not always sufficiently flexible to model unoccupied states. This is particularly
true for atoms and small dimers, where the LUMO has a character that is very different
from a linear combination of atomic like orbitals. 
In most cases, these basis set issues lead to small but noticeable errors on the level of DFT, 
but they are dramatically amplified at the level of $G_0W_0$.
For Xe,  H$_2$O and CH$_4$, GTO calculations with improved basis sets have been reported
finding very good to good agreement with 
the PW results. 

If we disregard the slightly disconcerting propagation of errors in going from DFT to $G_0W_0$ for LUMOs, 
we are satisfied by the  agreement between plane waves and
Gaussian type orbitals. As already stated, for the HOMO 
the mean absolute deviation is only 60 meV, which is excellent if one considers  that the computational
details are so different. Furthermore, our results have been
obtained using the $GW$ PAW potentials distributed with vasp.5.4,
so that similar calculations e.g. for molecules adsorbed on surfaces 
can be readily performed using the projector augmented wave method.

\begin{acknowledgments}
This work was supported by the Austrian Science Fund (FWF) within the
\emph{Spezialforschungsbereich Vienna Computational Materials Laboratory} (SFB
ViCoM, F41) and the \emph{Deutsche Forschungsgruppe Research Unit} FOR 1346.
P. Liu is grateful to the China Scholarship Council (CSC)-FWF Scholarship Program.
Computational resources were provided by the Vienna Scientific Cluster (VSC) and 
supercomputing facilities of the Universit\'{e} catholique de Louvain (CISM/UCL).
\end{acknowledgments}

\bibliography{master,GW}
\end{document}